\documentclass[usenatbib,fleqn,useAMS]{mnras}

\usepackage{amsmath}
\usepackage{mathrsfs}
\usepackage[varg]{txfonts}
\usepackage{braket}
\usepackage{cancel}

\renewcommand\le\oldleq
\renewcommand\ge\oldgeq

\newcommand{\dotp}{\bmath\cdot}
\newcommand{\Real}{\mathbb R}
\newcommand\dm{\mathrm d}
\newcommand{\pdm}{\upartial}
\renewcommand\pi\upi

\newtheorem{theorem}{Theorem}

\defcitealias{GNN1}{GNN}

\title[Symmetries of stellar systems]
{Reflection symmetries of Isolated Self-consistent Stellar Systems}

\author[An, Evans \& Sanders]
{J.~An$^1$\thanks{E-mail:~jinan@nao.cas.cn;~nwe,jls@ast.cam.ac.uk},
N.~W.~Evans$^2$, and J.~L.~Sanders$^2$
\\$^1$National Astronomical Observatories,
Chinese Academy of Sciences, A20 Datun Road, Chaoyang District,
Beijing 100012, China;
\\$^2$Institute of Astronomy, University of Cambridge, Madingley Road,
Cambridge CB3~0HA.}

\pagerange{\pageref{firstpage}--\pageref{lastpage}}\volume{000}\pubyear{2017}
\date{version \today.}

\begin{document}
\label{firstpage}
\maketitle

\begin{abstract}
Isolated, steady-state galaxies correspond to equilibrium solutions of
the Poisson--Vlasov system. We show that (i) all galaxies with a
distribution function (DF) depending on energy alone $f(E)$ must be
spherically symmetric and (ii) all axisymmetric galaxies with a DF
depending on energy and the angular momentum component parallel to the
symmetry axis $f(E,L_z)$ must also be reflection-symmetric about the
plane $z=0$. The former result is known, whilst the latter result is
new. These results are subsumed into the \emph{Symmetry Theorem},
which specifies how the symmetries of the DF in configuration or
velocity space can control the planes of reflection symmetries of the
ensuing stellar system.
\end{abstract}

\section{Introduction}

The shapes of isolated, steady-state stellar systems are controlled by
gravity. In such systems, the phase space distribution function (DF)
satisfies the collisionless Boltzmann equation (CBE) involving the
Newtonian potential, which is also coupled to the density (i.e.\ an
integrated DF) through Poisson's equation. This imposes severe
restrictions on the possible intrinsic shapes of systems. In fact, all
known equilibrium models of stellar systems are highly
symmetric. Spherically symmetric models were first studied by
J.~H.~Jeans and A.~S.~Eddington nearly a century ago. Algorithms to
find both isotropic and anisotropic DFs for spherical galaxies are now
well-established \citep[e.g.,][]{Ed16,Os79,Me85,De86,Ev06}. Methods to build
axisymmetric models with DFs depending on the two classical integrals
(energy $E$ and angular momentum component parallel to the symmetry
axis $L_z$) are also known \citep{LB62,Hu93}, together with some exact
solutions \citep{To82,Ev93}. Both spherically symmetric and
axisymmetric models contain an infinite number of reflectional planes
of symmetry. There are a very few triaxial models with DFs known
\citep{Va80,Hu92,Sa15}. However, even triaxial models have three
reflectional symmetries in the principal planes (the $D_{2h}$ point
group). This led \citet{Tr93} to raise the question as to whether
equilibrium models of stellar systems with still fewer symmetries can
exist.

For fluid dynamical equilibria, the principal result in this area was
established in the early years of the last century \citep[][see also
\citealt{Gr96} for a simplified treatment]{Li28}. Lichtenstein
studied self-gravitating, barotropic fluids and showed that if there
is a constant vector field $\hat\Bbbk$ such that the velocity field $\bmath v$
are stratified on the set of planes perpendicular to $\hat\Bbbk$ (viz.\ $\bmath
v\dotp\hat\Bbbk=0$ everywhere), then the figure has a symmetry plane
perpendicular to $\hat\Bbbk$. For a static fluid, there is a symmetry plane
perpendicular to every axis. Hence, all isolated, static,
self-gravitating, barotropic fluids must be spherical. The extension
of this result to stellar dynamics is given in \citet{BT}. These
authors pointed out that stellar dynamical models with ergodic DFs
$f(E)$ satisfy the self-same equations -- Poisson's equation,
hydrostatic equilibrium and density constant on equipotentials -- as
barotropic self-gravitating fluids. Thus, any isolated, static,
self-gravitating, ergodic stellar system must also be spherically
symmetric.

In this paper, we investigate whether one can generalize such
arguments to constrain the shape of relaxed stellar systems. In
Sect.~\ref{sec:oned}, we first examine the idealized one dimensional
case, which reduces the fundamental mathematical principle to the
level of elementary calculus. In Sect.~\ref{sec:GNN}, we introduce
some important mathematical results by \citet*[]{GNN1,GNN2}, which
essentially develop the one-dimensional analytical idea for higher
dimensions. As an illustration, we discuss how to recover the known
results on \citeauthor{Li28}'s theorem and ergodic systems using these
results \citep[cf.][]{PA96,Ci01}.  We then establish new results on
the reflection symmetry of systems built by the axisymmetric
two-integral DFs, $f(E,L_z)$ (Sect.~\ref{sec:symb}) as well as by DFs
satisfying certain sets of the symmetry conditions
(Sect.~\ref{sec:symc}).

\section{One dimensional case}
\label{sec:oned}

Let us consider the one-dimensional (henceforth ``1-d'')
stellar system in a steady-state
equilibrium with the potential $\Phi(x)$. The system is described by
the distribution function (DF) $F(x,v)$ that is a solution to the
collisionless Boltzmann equation (CBE), sometimes also called
the Vlasov equation;
\begin{equation}
\cancel{\frac{\pdm F}{\pdm t}}
+v\frac{\pdm F}{\pdm x}
-\frac{\dm\Phi}{\dm x}\frac{\pdm F}{\pdm v}=0.
\end{equation}
whose general solution is found via the method of characteristics to
be $F(x,v)=f(v^2/2+\Phi)$, where $f(E)$ is an arbitrary non-negative
function of $E$. In other words, any distribution satisfying the 1-d
time-independent CBE must be constant on the hyper-surfaces of
constant energy $E=v^2/2+\Phi(x)$ (Jeans' Theorem). The density $\rho$
of the system follows integrating the DF over the momentum
(i.e.\ velocity) space;
\begin{equation}\label{eq:rhoit}
\rho=\int_{-\infty}^\infty\!\dm v\,F(x,v)
=\!\sqrt2\!\int_\Phi^\infty\!\frac{f(E)\,\dm E}{\!\sqrt{E-\Phi}},
\end{equation}
which depends on the position only through the potential. That is to
say, the local density of the 1-d steady-state stellar system is
constant on locations with equal value of the potential and so the
density may be considered as a function of the potential,
$\rho=\rho(\Phi)$. If the potential $\Phi$ is generated
self-consistently by the density field $\rho(\Phi)$, then it must
satisfy the 1-d Poisson equation $\dm^2\Phi/\dm x^2=2G\rho(\Phi)$,
which results in an autonomous (i.e.\ not involving the independent
variable $x$ explicitly) second-order ordinary differential equation
$\Phi$.

We now show that, if there is a critical position in a 1-d potential,
then the potential and the density must be reflection-symmetric with
respect to the critical point. This result has been established before
by \citet{Sc13}, who however assumed existence of a critical point in
the potential implicitly (in fact, it is possible to have a solution
$\Phi$ that is strictly monotonic everywhere).  Here we rederive the
result somewhat more rigorously.  The proof may be constructed via
solving Poisson's equation formally for the solution, which is
achieved by reducing the degree of the differential equation. In
particular, if $N(\Phi)$ is the anti-derivative of $4G\rho$ considered
as a function of $\Phi$, that is, $\dm N(\Phi)/\dm\Phi=4G\rho$, then
Poisson's equation implies
\begin{equation}
\frac\dm{\dm x}\left[\left(\frac{\dm\Phi}{\dm x}\right)^2-N(\Phi)\right]
=2\left(\frac{\dm^2\Phi}{\dm x^2}-2G\rho\right)\frac{\dm\Phi}{\dm x}=0.
\end{equation}
In other word, $(\Phi')^2-N(\Phi)$ is constant for all positions
(within a connected interval over which $\Phi$ is finite).  Next
suppose that there exists $x_0$ such that $\Phi'(x_0)=0$.  It follows
from the constancy of $(\Phi')^2-N(\Phi)$ that the potential at any
location $x$ satisfies
\begin{equation}\label{eq:dp2}
\left(\frac{\dm\Phi}{\dm x}\right)^2=N(\Phi)-N(\Phi_0)
=4G\!\int_{\Phi_0}^\Phi\!\rho\,\dm\Phi=D(\Phi)\ge0,
\end{equation}
where $\Phi_0=\Phi(x_0)$, which should be the global minimum of $\Phi$
(i.e.\ $\Phi\ge\Phi_0$), provided that $\rho\ge0$ everywhere
(conversely, if $\rho\le0$ everywhere, then $\Phi\le\Phi_0$) in order
for $D(\Phi)$ to be non-negative. Equation (\ref{eq:dp2}) also
indicates that $|\dm x/\dm\Phi|=D^{-1/2}$ and so it follows (assuming
$\rho\ge0$ and $\Phi\ge\Phi_0$) that
\begin{equation}\label{eq:ivp}
|x-x_0|=\int_{\Phi_0}^{\Phi(x)}\frac{\dm\Phi}{\!\sqrt{D(\Phi)}}.
\end{equation}
Unless $\rho(\Phi)=0$ in an open neighborhood of $\Phi_0=\Phi(x_0)$,
$D(\Phi)$ is strictly positive and non-decreasing for $\Phi>\Phi_0$.
It follows that the right-hand side of equation (\ref{eq:ivp}) is
also a monotonic increasing function of $\Phi\ge\Phi_0$. Inverting
equation (\ref{eq:ivp}) for $\Phi(x)$ as a function of $x$ provides
the solution to Poisson's equation with the initial condition
that $\Phi'(x_0)=0$ and $\Phi(x_0)=\Phi_0$. Equation (\ref{eq:ivp})
further suggests that $\Phi(x)$, given the initial
condition, only depends on the distance $|x-x_0|$ to the critical
point $x_0$, which is to say $\Phi(x)$ is symmetric under
the reflection about $x=x_0$.

\section{The Gidas--Ni--Nirenberg theorems}
\label{sec:GNN}

Although the 1-d case elucidates the basic principle,
applying this analytic result to three dimensional problems requires
considerable mathematical finesse. Instead, we note
well-established results that generalize the discussion in
the preceding section for multi-dimensional spaces.

The results by \citet*{GNN1,GNN2} are particularly notable in this
respect. Whilst their results are celebrated amongst those who study
elliptic partial differential equations, they appear not widely known
in the astrophysical community. The most basic result of
\citet[henceforth GNN]{GNN1} is that
\begin{theorem}[GNN1]
If $\psi(\bmath x)$ is a positive $C^2$-function on the closed ball
of radius $R$ around the origin in $\Real^n$ satisfying
\begin{displaymath}\begin{cases}
\nabla^2\psi+f(\psi)=0&(\lVert\bmath x\rVert<R),\\
\psi(\bmath x)=0&(\lVert\bmath x\rVert=R),
\end{cases}\end{displaymath}
where $f(\psi)$ is a $C^1$-function of $\psi$,
then $\psi$ is radially symmetric and $\pdm\psi/\pdm r<0$
for $0<r=\lVert\bmath x\rVert<R$.
\end{theorem}
Extending to the whole space:
\begin{theorem}[GNN4]
Let $\psi(\bmath x)$ be a positive $C^2$-solution of
\begin{displaymath}
\nabla^2\psi+f(\psi)=0\quad\text{in $\Real^n$}
\end{displaymath}
with a $C^1$-function $f(\psi)$.
If $\psi$ admits the asymptotic expansion (with a fixed $m>0$)
up to a translation,
\begin{displaymath}
\psi=\frac{a_0}{\lVert\bmath x\rVert^m}
+\frac{\sum_{i,j}a_{ij}x_ix_j}{\lVert\bmath x\rVert^{m+4}}
+o\left(\frac1{\lVert\bmath x\rVert^{m+2}}\right)\quad
(\lVert\bmath x\rVert\to\infty)
\end{displaymath}
where $\bmath x=(x_1,\dotsc,x_n)\in\Real^n$, then
$\psi$ is radially symmetric and $\pdm\psi/\pdm r<0$
(where $r$ is the radial coordinate).
\end{theorem}
This further generalizes:
\begin{theorem}[GNN4']
If $\psi$ is a positive $C^2$-solution of
\begin{displaymath}
\nabla^2\psi+F(x_2,\dotsc,x_n;\psi)=0\quad\text{in $\Real^n$}
\end{displaymath}
with a continuous $\pdm F/\pdm\psi$, and expressible
in the same asymptotic series as the preceding theorem, then $\psi$
is symmetric under $\psi(-x_1,x_2,\dotsc,x_n)=\psi(x_1,x_2,\dotsc,x_n)$
and $\pdm\psi/\pdm x_1<0$ for $x_1>0$.
\end{theorem}
The theorems proved in \citetalias{GNN1} are of greater generality,
although we have here specialized to the specific case of Poisson's
equation.  The proofs are examples of the so-called moving plane
method, which relies on the maximum principle for the solution to some
classes of the elliptic partial differential equations.  An accessible
introduction for readers interested in the mathematical details of the
\citetalias{GNN1} Theorems is provided by the book of \citet{Fr00}. In
this paper, we will not attempt to reproduce these proofs, but accept
them as established facts to be applied. Nevertheless, we note that
the differentiability and the asymptotic behaviour conditions
appearing in the statements of the theorems (or the version of
theorems with appropriately relaxed conditions) are typically
satisfied by potentials due to physical models (characterized by
finite spatial extents or finite total mass with continuous and
bounded force field).

\subsection{Lichtenstein's Theorem in Fluid Mechanics}

Partly for pedagogical reasons, here we outline how the
\citetalias{GNN1} theorems lead to Lichtenstein's theorem in fluid
mechanics. \citeauthor{Li28}'s original paper (which predates
\citetalias{GNN1} by about a half century) is both lengthy and
somewhat inaccessible. Textbooks normally content themselves with
either stating the theorem \citep{Ta78,BT} or giving a simplified
proof for homogeneous fluids \citep{Gr96}. The modern
proof \citep[see e.g.,][sect.~4]{Li92}
employs similar techniques as \citetalias{GNN1}
(based on the maximum principle), but still obscures the connection to
the more general results.

Henceforth, we restrict ourselves to three dimensional space. In
addition, we also revert to the physics sign convention for the
potential: namely, that the acceleration is directed along the
downhill direction of the potential, $\bmath g=-\bmath\nabla\Phi$, which
results in the Poisson equation of $\nabla^2\Phi=4\pi G\rho$. For a
non-negative density $\rho\ge0$, the usual zero-point $\lim_{\lVert\bmath
r\rVert\to\infty}\Phi(\bmath r)=0$ then implies that the potential is
actually negative. Lichtenstein's Theorem states:
\begin{theorem}[Lichtenstein]
A time-independent barotropic fluid solution to the coupled
Euler--Poisson equation
\begin{displaymath}
\rho(\bmath v\dotp\bmath\nabla)\bmath v+\bmath\nabla p+\rho\bmath\nabla\Phi=0
;\quad\nabla^2\Phi=4\pi G\rho,
\end{displaymath}
with a stratified velocity field such that $\bmath v\dotp\hat\Bbbk=0$
(where $\hat\Bbbk$ is a fixed unit vector) is symmetric with respect
to the reflection about a plane perpendicular to $\hat\Bbbk$.
Furthermore, if the fluid is in a static equilibrium,
the system must also be spherically symmetric about the centre of mass
(and reflection symmetric with respect to any plane passing the centre).
\end{theorem}
Although this result predates the GNN theorems, Lichtenstein's theorem
is essentially a corollary. Following the barotropic assumption, it is
possible to define ``specific enthalpy'';
\begin{equation}
h(p)=\int_0^p\frac{\dm\tilde p}{\rho(\tilde p)},
\quad
\bmath\nabla h=\frac{\bmath\nabla p}\rho.
\end{equation}
Then, the Euler equation is reducible to
\begin{equation}
(\bmath v\dotp\bmath\nabla)\bmath v+\bmath\nabla(h+\Phi)=0,
\end{equation}
and the dot product to the fixed vector $\hat\Bbbk$ in the Cartesian
$z$-coordinate direction results in $(\pdm/\pdm z)(h+\Phi)=0$;
that is, $h+\Phi=C(x,y)$ is independent of the coordinate $z$ where
$C(x,y)$ is an arbitrary function of the coordinate components $(x,y)$
on the plane perpendicular to $\hat\Bbbk$. Applying the Laplacian on
$h+\Phi=C$ and using Poisson's equation then yields
\begin{equation}\label{eq:deqh}
\nabla^2h+4\pi G\rho(p)-\nabla^2C(x,y)=0.
\end{equation}
Since $\rho\ge0$, the enthalpy is a positive increasing function
of the pressure. Thus, $h(p)$ is in principle invertible for the pressure
as a function of the enthalpy, $p=p(h)$, and the barotropic density
can also be considered as a function of the enthalpy, $\rho(p(h))$.
Then equation (\ref{eq:deqh}) is in the form of $\nabla^2h+f_1(x,y;h)=0$
and the GNN theorem implies that the solution $h(\bmath r)$ is
reflection-symmetric, $h(x,y,-z)=h(x,y,z)$ with respect to a properly chosen
mid-plane. Since the pressure and the density are functions of the enthalpy,
they are also symmetric under the same reflection.
Alternatively, Poisson's equation directly indicates
$\nabla^2\Phi=4\pi G\rho(p)$ but $p=p(h)=p[C(x,y)-\Phi]$; that is,
$\nabla^2\psi+f_2(x,y;\psi)=0$ with $f_2=4\pi G\rho[p(C+\psi)]$
where $\psi=-\Phi$. The reflection symmetry of $\Phi$ is then
the result of GNN, whilst those of $h,p,\rho$ follow $h=C-\Phi$.

Spherical symmetry is an immediate corollary to the reflection symmetry,
since $\bmath v\dotp\hat\Bbbk=0$ for any $\hat\Bbbk$ in the static system.
However the barotropic assumption in a static equilibrium is actually
redundant, as the barotropy automatically follows the static Euler
equation; i.e.\ $\bmath\nabla\times(\rho^{-1}\bmath\nabla p+\bmath\nabla\Phi)
=-\rho^2(\bmath\nabla\rho\times\bmath\nabla p)=0\Rightarrow
\bmath\nabla\rho\parallel\bmath\nabla p$ (or $\bmath\nabla\rho=0$).
In fact, $\bmath\nabla\times(\bmath\nabla p+\rho\bmath\nabla\Phi)
=\bmath\nabla\rho\times\bmath\nabla\Phi=0$ by itself implies
$\bmath\nabla\rho\parallel\bmath\nabla\Phi$ and $\rho=\rho(\Phi)$;
that is, the isolated fluid system in a self-gravitating static equilibrium
must be spherically symmetric thanks to \citetalias{GNN1}.

\subsection{Ergodic Distribution Functions in Stellar Dynamics}

The \citetalias{GNN1} theorems also generalize the symmetry theorem of
the 1-d system proved in Sect.~\ref{sec:oned}. In the 1-d case, the
general solution to the time-independent CBE is an arbitrary function
of the specific energy. The Jeans theorem generalizes this for a three
dimensional system; that is, if $(I_1,I_2,I_3)$ is the set of three
independent isolating integrals of motion (admitted by the
one-particle Hamiltonian), any the DF (comprised of $N$ identical
particles governed by the same one-particle Hamiltonian) in
equilibrium must be constant over the joint level surfaces of
$(I_1,I_2,I_3)$ in phase space. This is usually stated as the DF is a
function of the integrals, $F(\bmath r,\bmath v)=f(I_1,I_2,I_3)$.

Since any time-independent Hamiltonian is itself an integral of
motion, the simplest DF in equilibrium is of the form $f(\mathcal H)$.
The Hamiltonian of a free particle in a fixed potential $\Phi$ is
$\mathcal H=\lVert\bmath v\rVert^2/2+\Phi(\bmath r)$.
As the Hamiltonian is isotropic (depending only on
the magnitude $v=\lVert\bmath v\rVert$),
the local density resulting from the DF of $f(\mathcal H)$ is then
\begin{equation}
\rho=4\pi\int_0^\infty\!\dm v\,v^2F(\bmath r,\bmath v)
=2^\frac52\pi\!\int_\Phi^\infty\!\dm\mathcal H
\sqrt{\mathcal H-\Phi}f(\mathcal H),
\end{equation}
which again depends on position only through the potential,
$\rho=\rho(\Phi)$. Therefore, following GNN, we conclude:
\begin{theorem}
The stellar dynamical system specified by an ergodic DF,
$F(\bmath r,\bmath v)=f(\lVert\bmath v\rVert^2/2+\Phi)$, must be spherically
symmetric about the centre of mass, provided that the total mass of
the system is finite and the potential $\Phi$ has been generated
self-consistently without any external potential.
\end{theorem}
\citet[Box~4.1]{BT} established this result based on Lichtenstein's
theorem, after showing that an ergodic DF results in the equation of
hydrostatic equilibrium. However, the \citetalias{GNN1} theorems
directly imply the sphericity of an ergodic stellar dynamical system,
as has already been noted in astrophysical literatures
\citep[e.g.,][]{PA96,Ci01,RG03}.


\section{Axisymmetric Stellar Systems}
\label{sec:symb}

We now apply the results of \citetalias{GNN1} to axisymmetric stellar
systems to obtain new results. Note that the stress tensor in such
systems is anisotropic and there is no possibility of recourse to
fluid mechanics and Lichtenstein's theorem, as the scalar pressure is
not defined. The situation is rectified thanks to the results of
\citetalias{GNN1}, which are still applicable.

If the potential is axisymmetric as in $\Phi(R,z)$ where $(x,y,z)$ are
the rectangular coordinates and $R^2=x^2+y^2$, then the axial
component of the angular momentum,
$L_z=\bmath L\dotp\hat\Bbbk=xv_y-yv_x$ (where $\bmath L=\bmath
r\times\bmath v$ is the specific angular momentum and $\hat\Bbbk$ is
the unit vector in the Cartesian $z$-direction) is also an integral of
motion, and so the two-integral DF $f(E,L_z)$ satisfies the CBE.
Although, in most studies, an axisymmetric two-integral system is
usually assumed to be symmetric about the reflection with respect to
the mid-plane, the assumption appears not to have been explicitly
proven previously. 
\begin{theorem}\label{th:f2refsym}
Consider a stellar dynamical system in an axisymmetric (about the
$z$-axis) potential $\Phi$. Suppose the system is specified by the
two-integral DF, $F(\bmath r,\bmath v)=f(E,L_z)$ where $E$ is the specific
energy and $L_z$ is the $z$-component of the specific angular momentum.
If the total mass of the system is finite and the potential $\Phi$ has been
generated self-consistently without any external potential, the system
must be symmetric with respect to the reflection about the plane
passing through the centre of mass and perpendicular to the $z$-axis.
\end{theorem}
In terms of the velocity component projected onto the orthonormal
frame for the cylindrical coordinates $(R,\phi,z)$, that is,
$(v_R,v_\phi,v_z)=(\dot R,R\dot\phi,\dot z)$, we find
$E=(v_R^2+v_\phi^2+v_z^2)/2+\Phi$ and $L_z=R^2\dot\phi=Rv_\phi$. Then
the density due to the DF $f(E,L_z)$ is obtained by the integral (here
$v_\wp^2=v_R^2+v_z^2$)
\begin{equation}\label{eq:f2den}
\rho=2\pi\!
\iint\limits_{v_\wp\ge0}\!F(\bmath r,\bmath v)\,v_\wp\,\dm v_\wp\,\dm v_\phi
=\frac{2\pi}R\!
\iint\limits_{L_z^2\le2R^2(E-\Phi)}\!\dm E\,\dm L_zf(E,L_z),
\end{equation}
which is axisymmetric (i.e.\ independent of the azimuth $\phi$).
Equation (\ref{eq:f2den}) indicates that the
$z$-dependence of the density is only through the potential $\Phi$,
that is, $\rho=\rho(R,\Phi)$. Then Poisson's equation for the
self-consistent system leads to the partial differential equation on
$\psi=-\Phi$ as in $\nabla^2\psi+4\pi
G\rho[\!\sqrt{x^2+y^2},-\psi]=0$, which is the form considered in the
theorem of GNN4'. Therefore, the potential $\Phi$ is reflection
symmetric with respect to a plane perpendicular to the $z$-axis (which
may be considered as the $z=0$ mid-plane without loss of generality)
and the conclusion of Theorem \ref{th:f2refsym} thus holds. In
particular, the symmetry of the density immediately follows the
symmetry of the potential,
$\rho(R,-z)=\rho[R,\Phi(R,-z)]=\rho[R,\Phi(R,z)]=\rho(R,z)$ (and so
the centre of mass consequently lies on the mid-plane), whilst the
symmetry of the velocity distribution is a simple consequence of the
symmetry of both $E$ and $L_z$. So if the two-integral DF exists,
  it is reflection-symmetric.

\section{Symmetric velocity distributions}
\label{sec:symc}

Let us rearrange the CBE in rectangular coordinates as
\begin{equation}\label{eq:cbe}
v_x\frac{\pdm F}{\pdm x}+v_y\frac{\pdm F}{\pdm y}
-\frac{\pdm\Phi}{\pdm x}\frac{\pdm F}{\pdm v_x}
-\frac{\pdm\Phi}{\pdm y}\frac{\pdm F}{\pdm v_y}
=\frac{\pdm\Phi}{\pdm z}\frac{\pdm F}{\pdm v_z}
-v_z\frac{\pdm F}{\pdm z},
\end{equation}
and consider the DF in equilibrium with both left and right hand sides
vanishing separately. The general solution for the vanishing right hand
side is found to be $F=f(v_z^2/2+\Phi;x,y;v_x,v_y)$ via the method of
characteristics \citep[e.g.,][]{Ga98} -- here
$v_z^2/2+\Phi=\mathcal H-(v_x^2+v_y^2)/2$ and so the DF is also of
the form $F=\tilde f(\mathcal H;x,y;v_x=p_x,v_y=p_y)$ albeit with a different
function $\tilde f$. The density resulting from this DF is found
from the integral (where $E_z=v_z^2/2+\Phi$)
\begin{equation}
\rho=\sqrt2\!\iiint\limits_{E_z\ge\Phi}\!\dm v_x\,\dm v_y\,\dm E_z
\frac{f(E_z;x,y;v_x,v_y)}{\!\sqrt{E_z-\Phi}},
\end{equation}
whose $z$-dependence is only through $\Phi$, namely,
$\rho=\rho(\Phi;x,y)$. This is again the form of the source term for
Poisson's equation assumed for the \citetalias{GNN1} theorems,
and so the potential $\Phi$ and the density $\rho$ are
reflection symmetric with respect to a plane perpendicular to the $z$-axis.
Since the DF is also an even function of $v_z$, the velocity distribution is
invariant under the same reflection too.

How can the DF satisfy the Cartesian $z$ part of the CBE separately
from the $(x,y)$ part? Let us observe that, with
$F=f(v_z^2/2+\Phi;x,y;v_x,v_y)$, equation (\ref{eq:cbe}) divides the
even and odd parts on $v_z$ to two opposite side. In general, an
arbitrary DF may be decomposed into
$F(x,y,z,;v_x,v_y,v_z)=F^+_{v_z}+F^-_{v_z}$ where $2F^\pm_{v_z}\equiv
F(x,y,z,;v_x,v_y,v_z)\pm F(x,y,z,;v_x,v_y,-v_z)$. The CBE for the
even-odd decomposed DF is then itself reassembled separately for the
even and odd parts, which results in
\begin{multline}
\frac{\pdm\Phi}{\pdm z}\frac{\pdm F^\pm_{v_z}}{\pdm v_z}
-v_z\frac{\pdm F^\pm_{v_z}}{\pdm z}
\\=\left(v_x\frac\pdm{\pdm x}+v_y\frac\pdm{\pdm y}
-\frac{\pdm\Phi}{\pdm x}\frac\pdm{\pdm v_x}
-\frac{\pdm\Phi}{\pdm y}\frac\pdm{\pdm v_y}\right)F^\mp_{v_z}.
\end{multline}
Hence, if $F^-_{v_z}=0$, then $F=F^+_{v_z}=f(v_z^2/2+\Phi;x,y;v_x,v_y)$;
\begin{theorem}\label{th:vzsym}
The finite-mass self-consistent stellar dynamic system specified by
the steady-state DF such that
$F(x,y,z;v_x,v_y,-v_z)=F(x,y,z;v_x,v_y,v_z)$ must possess a plane of
reflection symmetry perpendicular to the $z$-axis -- i.e.\ there
exists $z_0$ such that $\Phi(x,y,2z_0-z)=\Phi(x,y,z)$ and
$\rho(x,y,2z_0-z)=\rho(x,y,z)$. The DF is also expressible as
$F=f(v_z^2/2+\Phi;x,y;v_x,v_y)=\tilde f(\mathcal H;x,y;v_x,v_y)$ and so the
whole system is also invariant under the reflection about the $z=z_0$
plane (i.e.\ $z\to 2z_0-z$ and $v_z\to -v_z$).
\end{theorem}
This actually supersedes Theorem \ref{th:f2refsym}, for $f(E,L_z)$ is
invariant under $v_z\to-v_z$ given that $L_z=xv_y-yv_x$ is independent
of $v_z$ and $E=(v_x^2+v_y^2+v_z^2)/2+\Phi$ is an even function of $v_z$.

In fact, similar even-odd splits of the CBE can be applied for the DFs
with alternative sets of symmetries. In particular, the most general
result is:
\begin{theorem}[The Symmetry Theorem]\label{th:frsym}
The same conclusion as Theorem \ref{th:vzsym} holds if the DF is subject
to any one of the alternative symmetries,
\begin{enumerate}
\item $F(x,y,z;v_x,v_y,-v_z)=F(x,y,z;v_x,v_y,v_z)$,
\item $F(x,y,z;-v_x,-v_y,v_z)=F(x,y,z;v_x,v_y,v_z)$,
\item $F(x,y,2z_0-z;v_x,v_y,v_z)=F(x,y,z;v_x,v_y,v_z)$, or
\item $F(2x_0-x,2y_0-y,z;v_x,v_y,v_z)=F(x,y,z;v_x,v_y,v_z)$,
\end{enumerate}
where $x_0,y_0,z_0$ are fixed constants.
\end{theorem}
Here, the even-odd splits based on the last two symmetry assumptions
actually require the accompanying symmetry of the potential as in
$\Phi(x,y,2z_0-z)=\Phi(x,y,z)$ or $\Phi(2x_0-x,2y_0-y,z)=\Phi(x,y,z)$.
This however is a natural consequence of the self-consistency
condition (once the symmetry of the density is established following
the integration of the DF over the momentum space) and redundant.

Condition 3 appears to be same as the conclusion, but the conclusion
is actually more restrictive. In fact, the condition 3 does
\emph{not} describe the proper reflection symmetry of the DF with
respect to $z=z_0$ plane since it does not involve the transformation
of the velocity field (i.e.\ the true reflection symmetry follows from
invariance under $z\to2z_0-z$ and $v_z\to-v_z$). On the other hand,
the conclusion of the theorem implies the reflection symmetry of the
density, the potential and the DF, plus the DF being an even function
of $v_z$.

Similarly, condition 4 is \emph{not} in fact the true antipodal
symmetry about the axis defined by $x=x_0$ and $y=y_0$ (i.e.\ the
invariance under the 180\degr-rotation around the same axis). Rather
the condition indicates that the density and the potential is
antipodally symmetric whilst the velocity distributions in the axially
antipodal points are invariant under the rigid translation (but not
necessarily under 180\degr-rotation). Consequently, neither the DF
with rectangular reflection symmetry nor the one with the axial
rotational symmetry satisfy condition 4 unless some additional
conditions are imposed on the behaviour of the velocity
distributions. One such condition for the axially symmetric DF would
be the isotropy within the $v_x$-$v_y$ plane as in
$F=f(R,z;v_x^2+v_y^2,v_z)$, whereas the DF given by
$F=f(x^2,y^2,z;v_x^2,v_y^2,v_z)$ is symmetric under the individual
reflections, $x\to-x$ and $y\to-y$ and also satisfies the condition 4.

\subsection{Properties of Systems satisfying the Symmetry Theorem}

Since the DF in the form of $F=f(v_z^2+\Phi;x,y;v_x,v_y)$ is
an even function of $v_z$, any velocity moment with an odd power
to $v_z$ for this system vanishes. According to \citet[corollary~8]{AE16},
this implies the potential is separable like
$\Phi(x,y,z)=\Phi_1(x,y)+\Phi_2(z)$
unless $(\langle v_x^2\rangle-\langle v_z^2\rangle)
(\langle v_y^2\rangle-\langle v_z^2\rangle)=\langle v_xv_y\rangle^2$.
However, such separable potentials give rise
to self-consistent density profiles like $\rho=\rho_1(x,y)+\rho_2(z)$,
which cannot be of a finite total mass (except for $\rho=0$).

Consequently, the DFs of Theorem \ref{th:vzsym} or \ref{th:frsym}
must also be constrained so that
$(\langle v_x^2\rangle-\langle v_z^2\rangle)
(\langle v_y^2\rangle-\langle v_z^2\rangle)=\langle v_xv_y\rangle^2$.
Since $\langle v_xv_z\rangle=\langle v_yv_z\rangle=0$,
the characteristic polynomial of the matrix
resulting from the stress tensor is $(\langle v_z^2\rangle-\lambda)
[(\langle v_x^2\rangle-\lambda)(\langle v_y^2\rangle-\lambda)
-\langle v_xv_y\rangle^2]=0$; that is, $\lambda=\langle v_z^2\rangle$ is
one of the eigenvalues, whereas the constraints
$(\langle v_x^2\rangle-\langle v_z^2\rangle)
(\langle v_y^2\rangle-\langle v_z^2\rangle)=\langle v_xv_y\rangle^2$
is equivalent to $\langle v_z^2\rangle$ being one of the two remaining
eigenvalues. In other words, the velocity ellipsoids must be
spheroidal with its unequal axis aligned within the $x$-$y$ plane or
spherical everywhere. The two-integral distribution
$F=f(E,L_z)$ is an example of a DF satisfying such a constraint.

There are also restrictions on the potential.  In particular, the
vanishing left-hand side of equation (\ref{eq:cbe}) results in the
additional partial differential equation
\begin{equation}
v_x\frac{\pdm F}{\pdm x}+v_y\frac{\pdm F}{\pdm y}
=\frac{\pdm\Phi}{\pdm x}\frac{\pdm F}{\pdm v_x}
+\frac{\pdm\Phi}{\pdm y}\frac{\pdm F}{\pdm v_y}.
\end{equation}
Next, the $z$-derivative of this with the substitution
$v_z(\pdm F/\pdm z)
=(\pdm\Phi/\pdm z)(\pdm F/\pdm v_z)$ results in
\begin{equation}\label{eq:pde}
\frac{\pdm^2\Phi}{\pdm x\pdm z}
\left(v_x\frac{\pdm F}{\pdm v_z}-v_z\frac{\pdm F}{\pdm v_x}\right)
=\frac{\pdm^2\Phi}{\pdm y\pdm z}
\left(v_z\frac{\pdm F}{\pdm v_y}
-v_y\frac{\pdm F}{\pdm v_z}\right).
\end{equation}
However, the DF satisfying $v_z(\pdm F/\pdm z)
=(\pdm\Phi/\pdm z)(\pdm F/\pdm v_z)$ is also
expressible as $F=\tilde f(\mathcal H;x,y;v_x,v_y)$, and the partial
derivatives of $F$ with respect to the phase space coordinate
$(x,y,z;v_x,v_y,v_z)$ are related to the partial derivatives of
$\tilde f$ via
\begin{equation}\begin{split}
\frac{\pdm F}{\pdm x}
&=\frac{\pdm\Phi}{\pdm x}\frac{\pdm\tilde f}{\pdm\mathcal H}
+\frac{\pdm\tilde f}{\pdm x};\quad
\frac{\pdm F}{\pdm v_x}
=v_x\frac{\pdm\tilde f}{\pdm\mathcal H}
+\frac{\pdm\tilde f}{\pdm v_x},
\\
\frac{\pdm F}{\pdm z}
&=\frac{\pdm\Phi}{\pdm z}\frac{\pdm\tilde f}{\pdm\mathcal H}
;\quad
\frac{\pdm F}{\pdm v_z}
=v_z\frac{\pdm\tilde f}{\pdm\mathcal H}.
\end{split}\end{equation}
Thus equation (\ref{eq:pde}) reduces to a differential equation on
$\tilde f$;
\begin{equation}\label{eq:pde2}
\frac{\pdm^2\Phi}{\pdm x\pdm z}\frac{\pdm\tilde f}{\pdm v_x}
+\frac{\pdm^2\Phi}{\pdm y\pdm z}
\frac{\pdm\tilde f}{\pdm v_y}=0.
\end{equation}
Here the second-order derivatives on $\Phi$ cannot be identically
zero.\footnote{If otherwise, the potential is separable, leading to an
infinite mass.} Then the general solution of equation (\ref{eq:pde2})
for $\tilde f$ at a fixed $(\mathcal H,x,y)$ follows the method of
characteristics; that is, $\tilde f=\tilde f(\mathcal H;x,y;av_x+bv_y)$
where $a,b$ should be functions of only $(x,y)$\footnote{Formally
$a,b$ can be functions of $(\mathcal H,x,y)$. However, $\Phi$ is a function
of $(x,y,z)$ whilst $z$ and $\mathcal H$ are independent of each other
at a fixed $(x,y)$.} that satisfies
\begin{equation}\label{eq:pot2}
a(x,y)\frac{\pdm^2\Phi}{\pdm x\pdm z}
+b(x,y)\frac{\pdm^2\Phi}{\pdm y\pdm z}=0.
\end{equation}
This is possible only if $[\pdm^2\Phi/(\pdm x\pdm z)]/
[\pdm^2\Phi/(\pdm y\pdm z)]$ is independent of $z$.
Equivalently, there exist functions $a,b,c,d$ of $(x,y)$ and $G(x,y,z)$
such that the potential is restricted to be
\begin{equation}\label{eq:pot1}\begin{split}
\frac{\pdm\Phi}{\pdm x}&=c(x,y)+b(x,y)G(x,y,z);
\\\frac{\pdm\Phi}{\pdm y}&=d(x,y)-a(x,y)G(x,y,z).
\end{split}\end{equation}
In general, if equations (\ref{eq:pot2}) and (\ref{eq:pot1}) hold, we
can introduce an orthonormal frame that is locally rotated by $\varphi$
relative to the rigid Cartesian frame such that the velocity components
projected onto the frame are $(\hat av_x+\hat bv_y,\hat av_y-\hat bv_x,v_z)$
where $\hat a=a/(a^2+b^2)^{1/2}=\cos\varphi$ and
$\hat b=b/(a^2+b^2)^{1/2}=\sin\varphi$.
This frame then corresponds to the set of eigenvectors for the stress
tensor resulting from the DF of $\tilde f=\tilde f(\mathcal H;x,y;av_x+bv_y)$
-- in particular, $\langle(\hat av_x+\hat bv_y)(\hat av_y-\hat bv_x)\rangle=0$.
Furthermore, the resulting velocity dispersions are also constrained that
$\langle(\hat av_y-\hat bv_x)^2\rangle=\langle v_z^2\rangle$.

For example, with an axisymmetric potential $\Phi=\Phi(R,z)$
where $R^2=x^2+y^2$, we find that
$\pdm\Phi/\pdm x=(x/R)(\pdm\Phi/\pdm R)$ and
$\pdm\Phi/\pdm y=(y/R)(\pdm\Phi/\pdm R)$.
Then $(a,b,c,d)=(-y,x,0,0)$ and $G=R^{-1}(\pdm\Phi/\pdm R)$,
whereas $[\pdm^2\Phi/(\pdm x\pdm z)]/
[\pdm^2\Phi/(\pdm y\pdm z)]=x/y$.
In fact, the general solution to equation (\ref{eq:pde2}) in this case
is then $\tilde f=\tilde f(\mathcal H,L_z;x,y)$ with $L_z=xv_y-yv_x$ but
the CBE now reduces to
$v_x(\pdm\tilde f/\pdm x)+v_y(\pdm\tilde f/\pdm y)=0$
whose general solution is $\tilde f=\tilde f(xv_y-yv_x)$ at a fixed
$(\mathcal H,L_z)$. In other words, the only possible DFs in equilibrium
with an axisymmetric but non-separable
-- i.e.\ $\pdm^2\Phi/(\pdm R\pdm z)\ne0$ --
potential subject to one of the conditions in
the Symmetry Theorem \ref{th:frsym} are the two-integral DFs of
$F=f(E,L_z)$ (including the ergodic DF as a special case).
As is well known, the resulting velocity ellipsoids of such DFs are
all axially aligned ($\langle v_Rv_\phi\rangle=0$)
with $\langle v_R^2\rangle=\langle v_z^2\rangle$.

\section{Conclusions}

There is an ample body of work on the reflection symmetries of isolated,
self-gravitating, equilibrium fluid systems -- or more colloquially,
stars. It is already known that all isolated equilibrium stellar
models have spatial symmetries \citep[and references therein]{Li92}.
In Newtonian gravity, all non-rotating spherical models must be
spherically symmetric. Also, all equilibrium stars must have a
reflection symmetry perpendicular to the rotation axis of the star.
Some of these results have even been extended to relativistic stars
\citep[see e.g.,][]{Li94}.

By contrast, the analogous problem in stellar dynamics has received
scant attention. Isolated, self-gravitating, equilibrium stellar
systems (or more colloquially, galaxies) must satisfy the Poisson and
collisionless Boltzmann equations (also known as the Poisson--Vlasov
system). This imposes strict requirements on the properties of the
solutions, and hence on the shapes of stellar dynamical equilibria. It
has long been known that isolated systems with ergodic distribution
functions $f(E)$ must be spherically symmetric. This was strongly
hinted by early works such as \citet{Ed15}, whilst \citet{BT} show how
it can be deduced from the famous theorem of \cite{Li28} in fluid
mechanics. Its relationship to the symmetry theorems of elliptic
differential equations \citep{GNN1,GNN2} has been also noted by
\citet{PA96} and \citet{Ci01}.

Here, we have used the \citet{GNN1} theorems to derive a number of new
results on axisymmetric systems (Theorems 6--8).  Specifically, we
have shown that axisymmetric two-integral distribution functions
$f(E,L_z)$ must give rise to stellar systems with a plane of
reflectional symmetry perpendicular to the symmetry axis (which can be
taken without loss of generality as the plane $z=0$). Although this is
often assumed, it does not appear to have been proven previously. We
have also stated a new theorem -- the Symmetry Theorem -- which gives
a set of sufficient conditions on the distribution function to ensure
that the model has an underlying plane of reflectional symmetry.
Strictly speaking, our theorems presuppose the existence of a solution
to the Poisson and collisionless Boltzmann equations, but if such a
solution exists, it must have the specified symmetries.

There are further outstanding open problems in this area, of which we
highlight two. First, although the Symmetry Theorem provides
sufficient conditions for the existence of a plane of reflectional
symmetry, it does not provide necessary conditions. This is
illustrated by the distribution functions of spheroidal St\"ackel
models~\citep[e.g.,][]{De88}. What are the necessary and sufficient
conditions for the existence of a plane of reflectional symmetry?
Second, as pointed out by \citet{Tr93}, all known isolated, static,
stellar dynamical equilibria are highly symmetric -- either spherical,
axisymmetric or triaxial. They have at least three planes of
reflectional symmetry. Do static, isolated, stellar dynamical
equilibria with fewer symmetries exist? We suspect that the answer to
this question is negative, but a solid proof is lacking.

Of course, the problem of the shapes of equilibrium models is not just
of academic interest. Many galaxies are clearly not in equilibrium,
being still shaped by violent merging and accretion events or ongoing
star-formation. However, there are isolated, steady-state galaxies
known, as well as galaxies in the voids of large-scale
structure \citep[e.g.,][]{Su06}. Their equilibrium shapes will
ultimately be controlled by the balance between gravity and motion
(or more accurately, gradient of the momentum flux), rather than the
effects of environment. It is to these lonely souls that our work is
directly applicable.

\section*{acknowledgments} This paper originated from the discussion during the
first author (JA)'s visit to the Institute of Astronomy (Cambridge), which
was supported in part by the Science and Technology Facilities Council
(STFC, UK)'s Consolidated Grant award to University of Cambridge.
Work by JA is supported by the Chinese Academy of Sciences (CAS)
Fellowships for Young International Scientists (Grant No.:2009Y2AJ7)
and also grants from the National Science Foundation of China (NSFC),
including ``LAMOST and Galactic Dynamics (Grant No.:11390372)''.
JLS thanks the STFC for financial support.

\label{lastpage}
\end{document}